\documentclass[12pt]{article}
\usepackage{graphicx}
\usepackage{caption}
\usepackage{algorithm}
\usepackage{algpseudocode}
\usepackage{amsmath,amsfonts,amssymb, amsthm}
\usepackage{pgfplots,pgfplotstable, tikz}

\pgfplotsset{compat=newest}

\usetikzlibrary{calc}

\newcommand{\logLogSlopeTriangle}[5]
{
	
	\pgfplotsextra
	{
		\pgfkeysgetvalue{/pgfplots/xmin}{\xmin}
		\pgfkeysgetvalue{/pgfplots/xmax}{\xmax}
		\pgfkeysgetvalue{/pgfplots/ymin}{\ymin}
		\pgfkeysgetvalue{/pgfplots/ymax}{\ymax}
		
		\pgfmathsetmacro{\xArel}{#1}
		\pgfmathsetmacro{\yArel}{#3}
		\pgfmathsetmacro{\xBrel}{#1-#2}
		\pgfmathsetmacro{\yBrel}{\yArel}
		\pgfmathsetmacro{\xCrel}{\xArel}
		
		\pgfmathsetmacro{\lnxB}{\xmin*(1-(#1-#2))+\xmax*(#1-#2)} 
		\pgfmathsetmacro{\lnxA}{\xmin*(1-#1)+\xmax*#1} 
		\pgfmathsetmacro{\lnyA}{\ymin*(1-#3)+\ymax*#3} 
		\pgfmathsetmacro{\lnyC}{\lnyA+#4*(\lnxA-\lnxB)}
		\pgfmathsetmacro{\yCrel}{\lnyC-\ymin)/(\ymax-\ymin)} 
		
		\coordinate (A) at (rel axis cs:\xArel,\yArel);
		\coordinate (B) at (rel axis cs:\xBrel,\yBrel);
		\coordinate (C) at (rel axis cs:\xCrel,\yCrel);
		
		\draw[#5] (A)-- node[pos=0.5,anchor=north] {1} (B);
		\draw[#5] (B)-- (C);
		\draw[#5] (C)-- node[pos=0.5,anchor=west] {#4} (A);

	}
}

\title{Edge coloring in unstructured CFD codes}
\author{Andrew Giuliani \and Lilia Krivodonova}

\begin{document}
	\date{\vspace{-5ex}}
	\maketitle

\begin{abstract}
	We propose a way of preventing race conditions in the evaluation of the surface integral contribution in discontinuous Galerkin and finite volume flow solvers by coloring the edges (or faces) of the computational mesh.  In this work we use a partitioning algorithm that separates the edges of triangular elements into three groups and the faces of quadrangular and tetrahedral elements into four groups; we then extend this partitioning to adaptively refined, nonconforming meshes.  We use the ascribed coloring to reduce code memory requirements and optimize accessing the elemental data in memory.  This process reduces memory access latencies and speeds up computations on graphics processing units.
\end{abstract}

\maketitle

\vspace{-6pt}

\section{Introduction} \label{sec:intro}

Graphics processing units (GPUs) are massively parallel platforms that have become useful in computational fluid dynamics (CFD) solvers.  On such architectures, data are stored in shared memory and manipulated by processes that solve the problem in parallel.  If multiple processes, or threads, write simultaneously to the same memory location, a race condition is created.  Computations can then have unpredictable results in the sense that they become dependent on the order by which contentious memory locations are accessed \cite{nvidiablog}.  Therefore, care must be taken to develop algorithms that do not lead to race conditions.

A race condition can arise in the evaluation of the surface integral in discontinuous Galerkin (DG) and finite volume type numerical methods.  We will illustrate this issue on an example of the DG method used to discretize the conservation law

\begin{equation} \label{conservationlaw}
	\frac{d}{dt}\mathbf{u} + \nabla \cdot \mathbf{F}(\mathbf{u})  = 0,
\end{equation}
with the solution $\mathbf{u}(\mathbf{x}, t) = (u_1, u_2,..., u_M)^\intercal$, $(\mathbf{x},t)\in \Omega \times [0,T]$, and the flux function $\mathbf{F}(\mathbf{u})$.  We divide the domain $\Omega$ into a mesh of elements, e.g., triangles, quadrilaterals, tetrahedra for two and three-dimensional problems.  Discretizing \eqref{conservationlaw} on this mesh with the DG method \cite{giuliani} yields the scheme 
\begin{align}
	\label{eq:dg}
	\frac{d}{dt} \mathbf{c}_{i,j} = \int_{\Omega_i} \mathbf{F}(\mathbf{U}_i) \cdot \nabla v_{i,j} d\Omega_i
	- \sum_{q} \int_{I_{i,q}} v_{i,j} \mathbf{F}(\mathbf{U}_i, \mathbf{U}_{p_q}) \cdot \mathbf{n}_{i,q} dI_{i,q},
\end{align}
where the numerical solution $\mathbf{U}_i$ on the physical element $\Omega_i$ is approximated by a linear combination of $N_p$ orthonormal basis functions $v_{i,j}$, i.e. $\mathbf{U}_i = \sum^{N_p}_{j=1}\mathbf{c}_{i,j} v_{i,j}$ with $\mathbf{c}_{i,j} = [c^1_{i,j}, c^2_{i,j},\dots, c^M_{i,j}]^\intercal$ as the modal degrees of freedom (DOFs). The numerical flux $\mathbf{F}(\mathbf{U}_i,\mathbf{U}_{p_q})$ is computed on the surface shared by adjacent cells $\Omega_i$ and $\Omega_{p_q}$, and $\mathbf{n}_{i,q}$ is the outward facing normal on the $q$th surface of element $i$.  We refer to the elements that share the surface $e_k$ as the left and right elements of that surface.  Depending on the dimension of the problem, surfaces of the element $\Omega_i$ can geometrically be either edges (of, e.g., a triangle) or faces (of, e.g., a tetrahedron).  For simplicity of illustration, we will mostly discuss edges although the ideas are applicable to faces as well.  

\begin{figure}
	\centering
	\includegraphics[scale = 1]{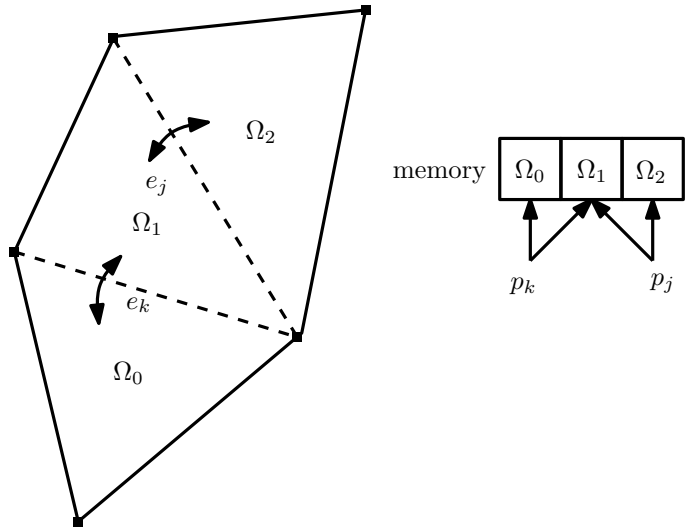}
	\caption{Sample mesh (left); processes $p_k$ and $p_{j}$ both write surface contributions to the memory location for $\Omega_1$ (right).}
	\label{fig:racecondition}
\end{figure}

Integrating the numerical solution in time requires evaluation of the right-hand side of \eqref{eq:dg}, which is composed of a volume integral over $\Omega_i$ and surface integrals over $I_{i,q}$.  The volume contribution is easy to parallelize as it only requires information local to $\Omega_i$.  However, the surface contribution requires more care as it involves writing data to memory locations for both $\Omega_i$ and its neighbors $\Omega_{p_k}$. The surface contributions can be computed in parallel by assigning one thread per edge in the computational mesh as shown in Figure \ref{fig:racecondition}.  The processes $p_k$ and $p_{j}$ are then tasked with computing the surface terms along edges $e_k$ and $e_j$, respectively.  Once this is done, the processes must save the result at memory addresses corresponding to the edge's left and right elements, e.g. $p_k$ writes the surface contribution to locations for $\Omega_0$, and $\Omega_1$, and $p_j$ writes to locations for $\Omega_1$, and $\Omega_2$. If both processes write simultaneously to the address for $\Omega_1$, a race condition will occur.

The mitigation of such a race condition can be done in a number of fashions.  First, each process can be given a portion of buffer memory such that no two processes store their result at the same location.  For example, process $j$ could write data for its left and right element to addresses $2j$ and $2j+1$, respectively.  Subsequently, these data in the buffer can be combined using an additional parallel kernel \cite{giuliani, nodaldg, corrigan}. The downside of such an approach is the extensive use of buffer memory for intermediate calculations.  The length of these buffer vectors is the number of DOFs per element times the number of surfaces; we report the memory requirements of these buffers in Section \ref{sec:memorysaved}.  Atomic operations have also been proposed, but may degrade the solver's efficiency by an order of magnitude \cite{giuliani}.  Finally, the element-wise surface integration approach has been discussed and shown to be suboptimal in \cite{giuliani,luo2, corrigan}, as fluxes at surface integration points are evaluated twice.

\begin{figure}
	\centering
	\includegraphics[trim=100 100 80 80, clip,width=\linewidth]{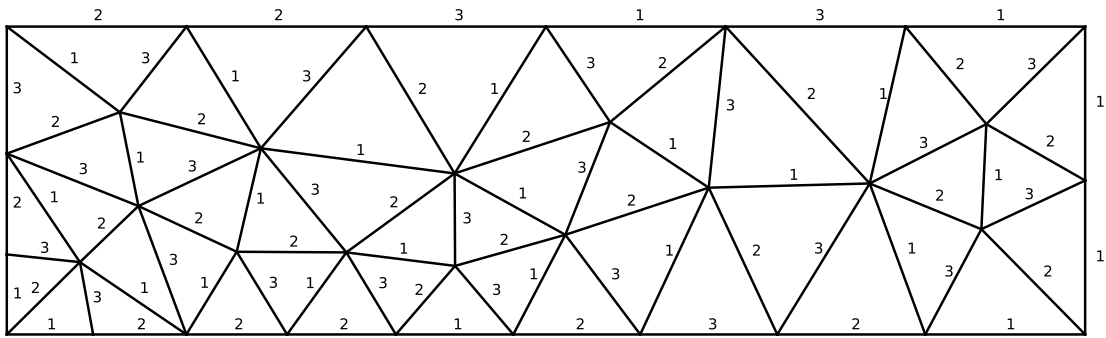}
	\caption{An example of edge coloring.  Each triangle has edges of distinct, non-repeated colors 1, 2 and 3.}
	\label{fig:mesh}
\end{figure}

Another approach to avoiding a race condition is coloring the surfaces of the mesh.  The surfaces of a mesh can be colored into separate groups such that no element possesses surfaces that belong to the same color.  A sample two-dimensional mesh and a possible edge coloring are shown in Figure \ref{fig:mesh}.  Each edge has a color (1,2, or 3) and no triangle has multiple edges of the same color.  Clearly, three is the minimum number of colors for this element geometry.  Once the set of edges is partitioned into three groups, the surface contributions are evaluated color-by-color and added to the variable $\mathbf{c^\text{rhs}}$, which stores the volume contribution as shown in Algorithm \ref{alg:fluxeval}.  
Because no edges in a color group share the same element, there will be no memory in contention, that is, the race condition will be avoided.

Edge coloring has been used in the context of MPI computing, for example \cite{lohner2008applied}, to distribute edges between cores.  It has also been used in GPU computing, e.g. \cite{luo,komatitsch}, where they were not concerned with using the optimal number of colors.  Typically, a naive greedy coloring algorithm will loop through all surfaces in the mesh.   For each surface, the edge colors of its the left and right element are checked.  Then, the first available color is assigned to the surface.  This greedy algorithm will yield meshes of a maximum five colors for triangular elements, and a maximum of seven colors for quadrilateral and tetrahedral elements.  This is not the minimum number of colors possible for these element geometries.  In this work, we propose an edge coloring heuristic that results in an optimal number of colors for a variety of element geometries.  This allows for a reduced number of kernel launches and streamlining of the code.  The overhead associated with launching extra kernels is minor for GPUs, but non-negligible for multi-core CPUs \cite{luo}.  Our testing on GPUs has shown that using a suboptimal number of colors can decrease execution efficiency by approximately 5 \%, which is negligible, see Table \ref{tab:timing}.  However, an excessive number of colors can be a detriment to code simplicity.  This is especially true with respect to adaptive mesh refinement, where the number of required colors can rapidly increase.  For example, with tetrahedral elements a greedy coloring algorithm would yield seven colors versus our algorithm which yields the optimal four colors.  Therefore, for code simplicity and a reduced number of kernel launches, it is worthwhile to use the fewest possible number of colors to partition the set of edges in the mesh.   

In order to improve the performance of unstructured CFD solvers, elements and sides may be ordered in memory to reduce memory access latencies, e.g., space-filling curves and the bin-ordering method have been used on single and multi-processor machines \cite{lohner2008applied, marsha,corrigan}.  The speed-up observed depends on the hardware architecture \cite{lohner2008applied} and software implementation.  Various renumbering techniques are compared in \cite{burgess} based on the performance of an edge-based solver for the Euler equations.   A maximum speed-up of approximately 20 \% was obtained compared to the ordered output from the mesh generator.  In this work, we propose an ordering scheme based on edge coloring that exhibits comparable speed-ups for GPUs and show that using a minimal number of colors maximizes the attainable speed-up.

Thus, to eliminate memory contention, reduce memory requirements, and speed-up computations, we color the surfaces of the mesh into separate groups, e.g. colors 1, 2, and 3 for triangular elements and 1,2,3, and 4 for tetrahedra.  The objective is to seek a coloring with the minimum number of colors.  First, we apply a modified greedy coloring algorithm described in Section \ref{sec:algorithm}.  This algorithm has linear complexity, but colors the mesh imperfectly, i.e., certain surfaces remain uncolored.  We then post-process the created greedy coloring to ensure all surfaces are colored with the minimum possible number of colors, also described in Section \ref{sec:algorithm}.  We numerically demonstrate that the coloring finishes in linear time on triangular, quadrilateral, and tetrahedral elements geometries.

\begin{algorithm}
	\caption{Surface contribution evaluation}  \label{alg:fluxeval}
	\begin{algorithmic}
		\Ensure the surface contribution stored in the variable $\mathbf{c^{rhs}}$. 
		\Procedure{surface evaluation}{}
		\For{$i \in \{1,2,...,N_\text{colors} \}$}
		\ForAll{$e_k$ in parallel}
		\State ($l$, $r$) $\gets$ the indices of two elements that share the edge $e_k$
		\For{$j \in [1,...,N_p]$}
		\State $\mathbf{c^{rhs}}_{l,j}\gets$ surface contribution for $\Omega_l$ along $e_k$
		\State $\mathbf{c^{rhs}}_{r,j}\gets$ surface contribution for $\Omega_r$ along $e_k$
		\EndFor
		\EndFor
		\State synchronize threads
		\EndFor
		
		\EndProcedure
	\end{algorithmic}  
\end{algorithm}

\section{Edge coloring}

The coloring of the edges (or faces) of the computational mesh can be related to a standard edge coloring problem from graph theory for the element-wise connectivity graph.  Given a computational mesh, we will now construct such a graph.

\begin{figure}
\centering
\includegraphics{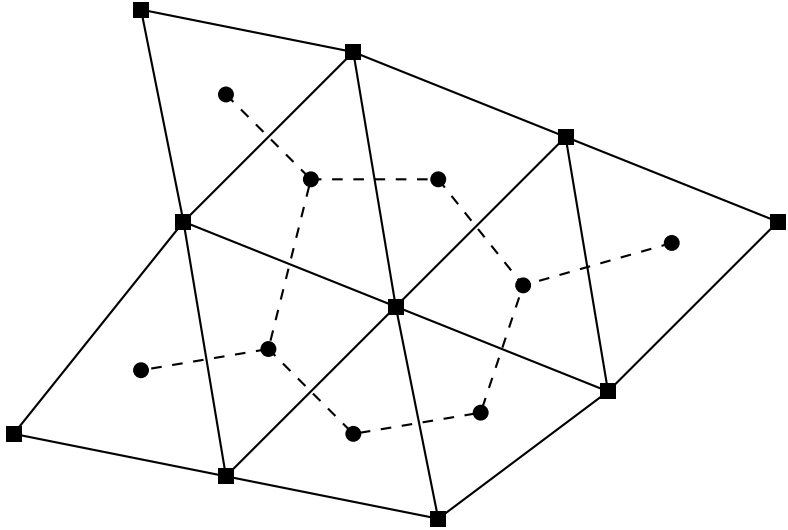}
\caption{A computational mesh is shown with solid edges and square vertices. The mesh's associated connectivity graph has dashed lines and circular nodes.}
\label{fig:graphandconnectivity}
\end{figure}

A computational mesh $(V,E,\Omega)$ can be described by its vertices $v_i \in V$, edges (or faces) $e_k \in E$, and elements $\Omega_i \in \Omega$.  For a given mesh, we can construct an associated graph $G(N,L)$, by first placing a node $n_i \in N$ at the barycenter of the element $\Omega_i \in \Omega$.  Further, two nodes $n_i, n_j$ are connected by a line $l_k \in L$ when the elements $\Omega_i$ and $\Omega_j$ are adjacent, i.e. they share an edge (or face).  In other words, $G$ is a graph showing the connectivity between elements in the computational mesh.  This process is illustrated in Figure \ref{fig:graphandconnectivity}.

The number of lines in $L$ incident on a node $n_i$ in $N$ is called the degree of $n_i$, $\deg(n_i)$.  Vizing's theorem \cite{vizing} states: a graph can be colored with $\max_{n_i \in N} \deg(n_i) + 1$ colors.  That is, the connectivity graph of meshes with triangular elements can always be colored with four colors.  Likewise, meshes with quadrilateral and tetrahedral elements can always be colored with five colors.

Furthermore, for planar meshes of triangles, it has been shown that only three colors are necessary \cite{wilson1996introduction}.  The proof is based on assigning a color to each geometrical vertex $v_i \in V$ of the planar mesh of triangles.  The colors must be placed such that no two vertices connected by a geometrical edge have the same color.  By the four color theorem \cite{appel}, this can be done with four colors if the mesh is a planar graph.  From this vertex coloring, it can be shown that a valid edge coloring with three colors is always possible.  This result also holds for meshes of triangles embedded on a sphere.  Although it is not intuitive, such meshes are equivalent, i.e., isomorphic to planar graphs.

\subsection{Approaches to coloring}
The coloring algorithm resulting from the constructive proof of the four color theorem \cite{appel} is too complex for practical mesh coloring because it requires the handling of hundreds of reducible configurations.  However, there exist heuristic algorithms that are simple and have been shown to terminate.  The color exchange algorithm presented in \cite{complexcolors} is a simple probabilistic algorithm that terminates in polynomial time for general graphs.  It is composed of two stages: first the mesh is colored imperfectly with a greedy algorithm, next a conflict resolution stage is completed to obtain an acceptable partitioning.  During this last stage, the algorithm sometimes enters an unterminating loop.  The loop can be broken by swapping colors in a prescribed manner.  An alternative method is to restart the entire algorithm with a new initial greedy coloring in the hopes that a loop is not encountered again.  This is the conflicting vertex displacement (CVD) algorithm for edge coloring graphs presented in \cite{fiol2012}.  We propose an algorithm with a simpler resolution of unterminating loops than the one presented in \cite{complexcolors} that does not require recoloring the entire mesh like the one in \cite{fiol2012}.

\section{Algorithm} \label{sec:algorithm}
Triangular and tetrahedral elements have 3 edges and 4 faces, respectively.  Consequently, the smallest possible number of colors is three for triangles and four for tetrahedra.  We call the set of available colors $C$, with $C = \{1,2,3\}$ for triangles and $C = \{1,2,3,4\}$ for quadrangles and tetrahedra.  We begin by describing the notation that will be used in the following subsections.  The $k$th edge (or face) in the computational mesh is named $e_k$, and the $i$th element in the computational mesh is named $\Omega_i$.  The two elements that share $e_k$ are denoted by $\Omega_{k_l}$ and $\Omega_{k_r}$, i.e. the left and right element of edge $e_k$.  A nonconflicting coloring of an element $\Omega_i$ is such that the edge colors of $\Omega_i$ do not repeat. A nonconflicting coloring of $e_k$ is such that both $\Omega_{k_l}$ and $\Omega_{k_r}$ have nonconflicting colorings.

\subsection{Conforming meshes}

We now describe our coloring algorithm for conforming meshes.  It is composed of two stages: a modified greedy coloring procedure, followed by a conflict resolution step.  To simplify the illustration, we describe the algorithm for triangular meshes.  However, analagous arguments can be made for meshes of other element geometries, such as quadrangles, tetrahedra, etc.

\begin{figure}
	\centering
	\includegraphics[scale = 1]{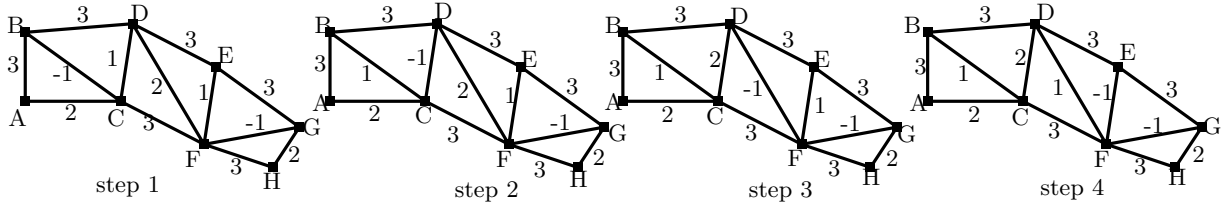}
	\caption{Both conflicts in step 1 can be resolved once the conflict originally on $BC$ reaches $EF$ in step 4.}
	\label{fig:resolve}
\end{figure}

At the start of the coloring algorithm, the color of each edge is initialized to -1, meaning that all edges have not been assigned a color.  Then, the modified greedy coloring procedure passes through all edges $e_k$ in the mesh.  For each edge $e_k$, the edge colors of its the left and right elements ($\Omega_{k_l}$ and $\Omega_{k_r}$) are checked.  
If possible, a nonconflicting color in $C$ is randomly assigned to $e_k$. A standard greedy coloring algorithm augments the set of colors when none are available to create a nonconflicting coloring for $\Omega_{k_l}$ and $\Omega_{k_r}$.  
Instead of increasing the number of colors in $C$, our modified greedy coloring algorithm leaves edges uncolored, i.e. of color \texttt{-1}, when there are no colors in $C$ that yield a nonconflicting coloring.  
For example, in step 1 of Figure \ref{fig:resolve}, edge $CB$ cannot be assigned a color without causing either triangle $ABC$ or $CBD$ to have a conflicting coloring.  
Uncolored edges are called conflicts. 

We now describe the procedure that seeks to resolve the conflicts created by the modified greedy algorithm, with the aim to obtain a nonconflicting edge coloring for all edges.  
We consider the uncolored edges one at a time.  
For an uncolored edge $e_k$, we exchange its color, i.e., \texttt{-1}, with the color of an edge that belongs to $\Omega_{k_l}$ or $\Omega_{k_r}$ such that the number of conflicts does not increase.  For example, in Figure \ref{fig:resolve}: step 1, the conflict on $CB$ may be swapped with the color of either edge $CA$ or $CD$.  
This operation does not increase the total number of conflicts in the mesh.  However, swapping colors of $BA$ and $BC$ will create a new conflict as $BCD$ will have two edges with the same color, 3.  
 
A simple geometric consideration reveals that a swap that does not increase the conflict count is always possible.  
After the swap, the conflict can either be resolved by choosing a nonconflicting color or it can be propagated further.  
For example, the conflict can be moved along the sequence of edges $BC$-$CD$-$DF$-$EF$ as illustrated in Figure \ref{fig:resolve}: steps 1, 2, 3, and 4.  
After that, the conflict on edge $EF$ can be resolved by assigning it to color 2 (Figure \ref{fig:resolve}: step 4).  Conflicts generally cancel out one another, or resolve once they reach a boundary edge.
Once a conflict is resolved, we move on to the next uncolored edge, i.e. conflict, and try to resolve it.  

\begin{figure}
	\centering
	\includegraphics[scale = .85]{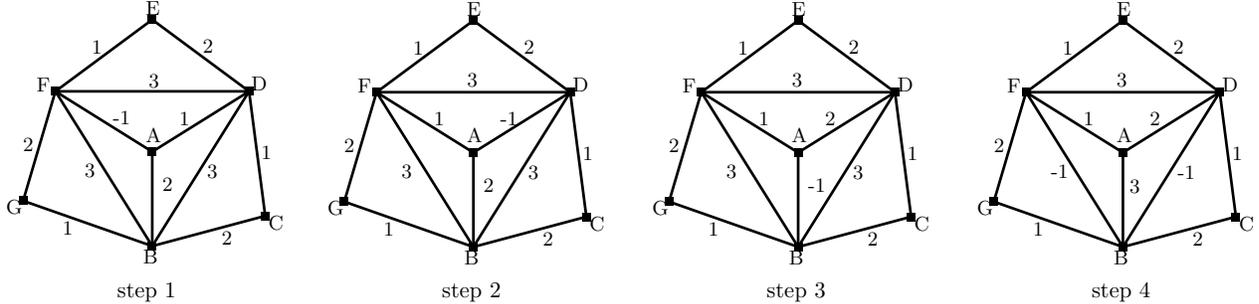}
	\caption{The conflict on $AF$ is stuck on a loop $AF$-$AD$-$AB$.  Moving the conflict from that loop to $FB$ creates an additional conflict on $DB$ in step 4.}
	\label{fig:cycle}
\end{figure}

Some conflicts will be resolved in a finite number of swaps, and some will create a loop and visit the same sequence of edges.  
The latter case is illustrated in Figure \ref{fig:cycle} steps 1, 2, and 3.  
A loop is detected when the conflict is directed back to an edge that has already been visited.  
In this case, the algorithm attempts to break the loop by choosing a swap that increases the number of conflicts by one.  For a conflict on edge $e_k$ that is in a loop, we uncolor two edges of $\Omega_{k_l}$ and $\Omega_{k_r}$ that share the same color, i.e. assign -1, to them.  
In step 4 of Figure \ref{fig:cycle}, the conflict in a loop on edge $AB$ is moved to both $BD$ and $BF$ (two edges that previously shared the same color).  
Thus, the number of conflicts in the mesh is increased by one.  
The algorithm then attempts to resolve the next conflicting edge in the mesh.

The greedy coloring runs linearly in the number of edges as each edge is only visited once.  We do not have a proof that the conflict resolution stage always terminates, so we cannot determine its theoretical complexity.  However, the runtime seems to scale linearly with the number of edges in the mesh according to the timings in Section \ref{sec:examples}.

\subsection{Nonconforming meshes}
In this section we propose a coloring algorithm for nonconforming meshes, which are often a result of adaptive mesh refinement.  We assume that computations start with a conforming mesh as most CFD codes do.  The set of edges in the original mesh is partitioned using the algorithm described in Section \ref{sec:algorithm}.  Based on this partition we color the refined mesh using the minimum possible number of colors.  This process can be performed naturally in parallel on a GPU.  The algorithm allows for easy transition between multiple levels of refinement without introducing conflicts.  In this section, we make an assumption that adjacent elements differ by at most one level of refinement, though the idea can be extended to less smooth meshes.  

\begin{figure}
	\centering
	\includegraphics[width = 0.7\linewidth]{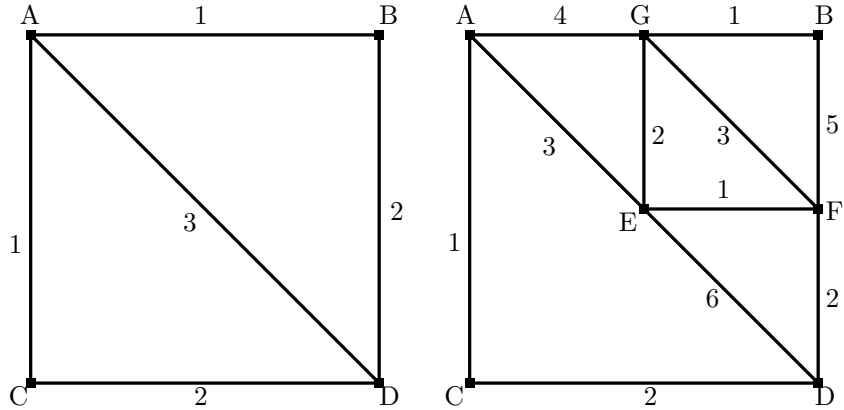}
	\caption{Original (left) and refined (right) meshes with colors.}
	\label{fig:isotropic}
\end{figure}

\begin{figure}
	\centering
	\includegraphics[width = 0.25\linewidth]{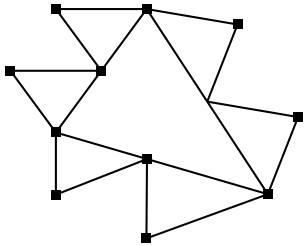}
	\caption{The maximum number of refined neighbors a coarse triangle (center) can have is six.}
	\label{fig:worstcase}
\end{figure}

One way a triangle can be refined is by splitting it into four smaller triangles by connecting the midpoints of its edges as shown in Figure \ref{fig:isotropic}.  This might produce a conconforming mesh if the adjacent element is not refined (Figure \ref{fig:isotropic}).  We propose a coloring of these smaller triangles that doubles the number of colors initially present in the mesh to 6.  This is the minimum possible number because a coarse element might have six refined neighbours (Figure \ref{fig:worstcase}).  We note that some refinement strategies will refine elements with six neighbors to improve mesh smoothness, but for completeness we not do so here \cite{dealii}.

We call the edges of the original (or parent) triangle parent edges, each with a parent color $c^{(n)}$ with $n = 1,2,3$.  The parent edges are ordered counterclockwise, e.g., the edges of triangle $DBA$ in Figure \ref{fig:isotropic}, left, are ordered $DB$, $BA$, and $AD$.  Therefore, in this example the parent colors are $c^{(1)} = 2$, $c^{(2)} = 1$ and $c^{(3)} =3$.

In Figure \ref{fig:isotropic}, right, triangle $DBA$ is refined.  Each parent edge is divided into two child edges of equal length, e.g. the second edge $BA$ becomes $BG$ and $GA$.  The first child inherits the color of the parent, and the second child's color is the parent's color shifted by three.  We can write this formally with the mapping $c_m^{(n)} = [(c^{(n)} + 3m -4) \text{ mod } 6]+1$ for the the $m$th child edge on the $n$th parent edge. The child edges are also ordered counterclockwise, e.g., the second parent edge $BA$ has child edges that are ordered $BG$, $GA$.  Therefore, in this example the color of each child edge is $c_1^{(2)} = 1$, $c_2^{(2)} = 4$. Further, three additional edges are created in the element's interior: $GE$, $EF$ and $FG$.  Each edge is prescribed the color of the parent edge to which it is parallel, e.g. $GF$ takes the color of $AD$ because they are parallel.

Coarsening the mesh consists of merging four small elements into one.  In the coarsened triangle, a parent edge takes the color of the child edge to which it is parallel.  That is, the coarsened triangle inherits the colors of the interior small triangle.  This approach recovers the coloring of the parent edges before refinement.

\section{Mesh coloring examples} \label{sec:examples}
In Tables \ref{tab:tri}, \ref{tab:quad} and \ref{tab:3D}, we present a number of examples on meshes of different sizes and geometries for the partitioning algorithm.  The coloring algorithm successfully assigns approximately the same number of edges to each available color in a reasonable amount of time.  All coloring examples were executed on an Intel Xeon CPU E5530 2.40GHz processor.

\begin{figure}
	\centering
  $\vcenter{\hbox{\includegraphics[width=0.45\linewidth]{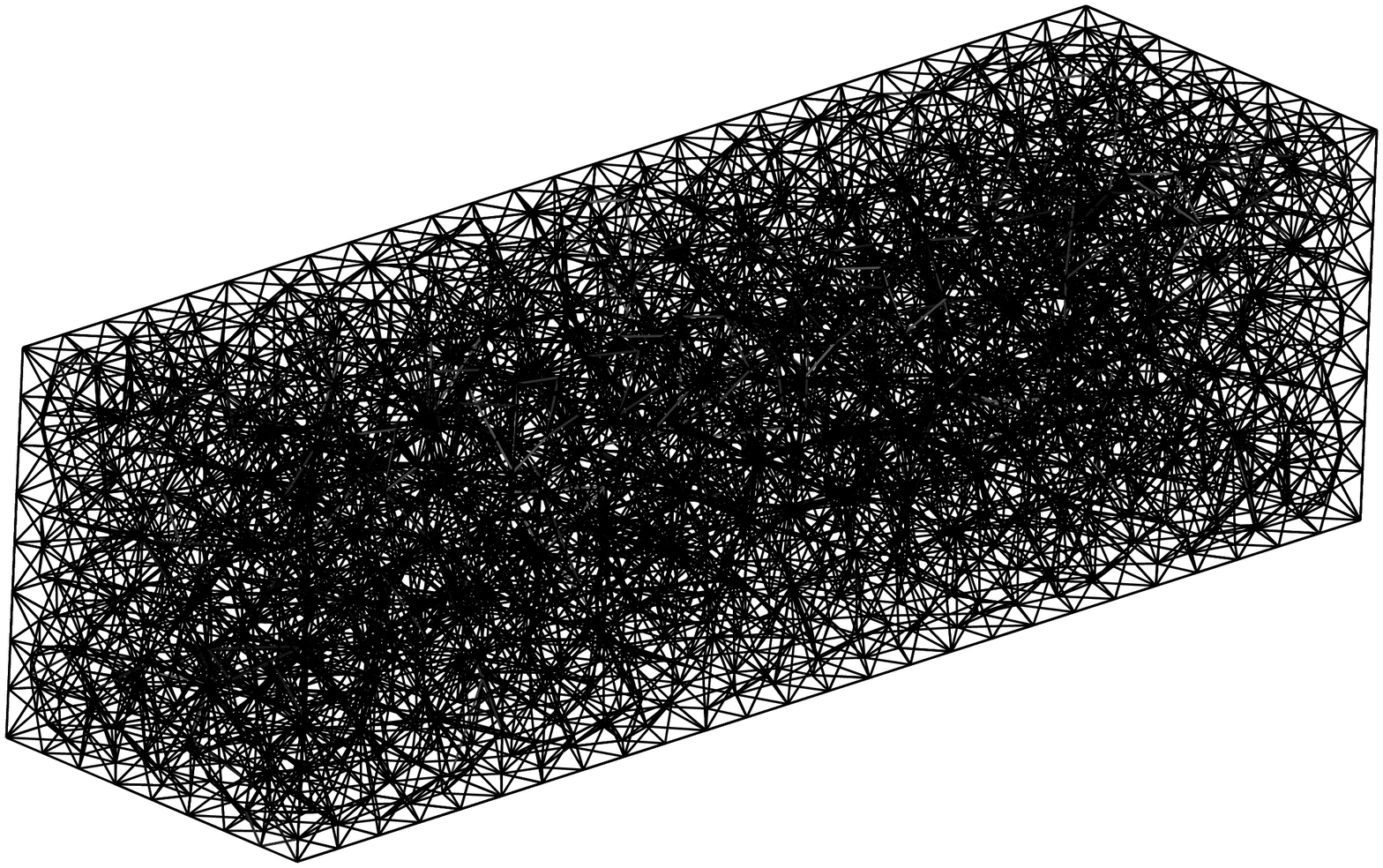}}}$
  \hspace*{.2in}
  $\vcenter{\hbox{\includegraphics[width=0.45\linewidth]{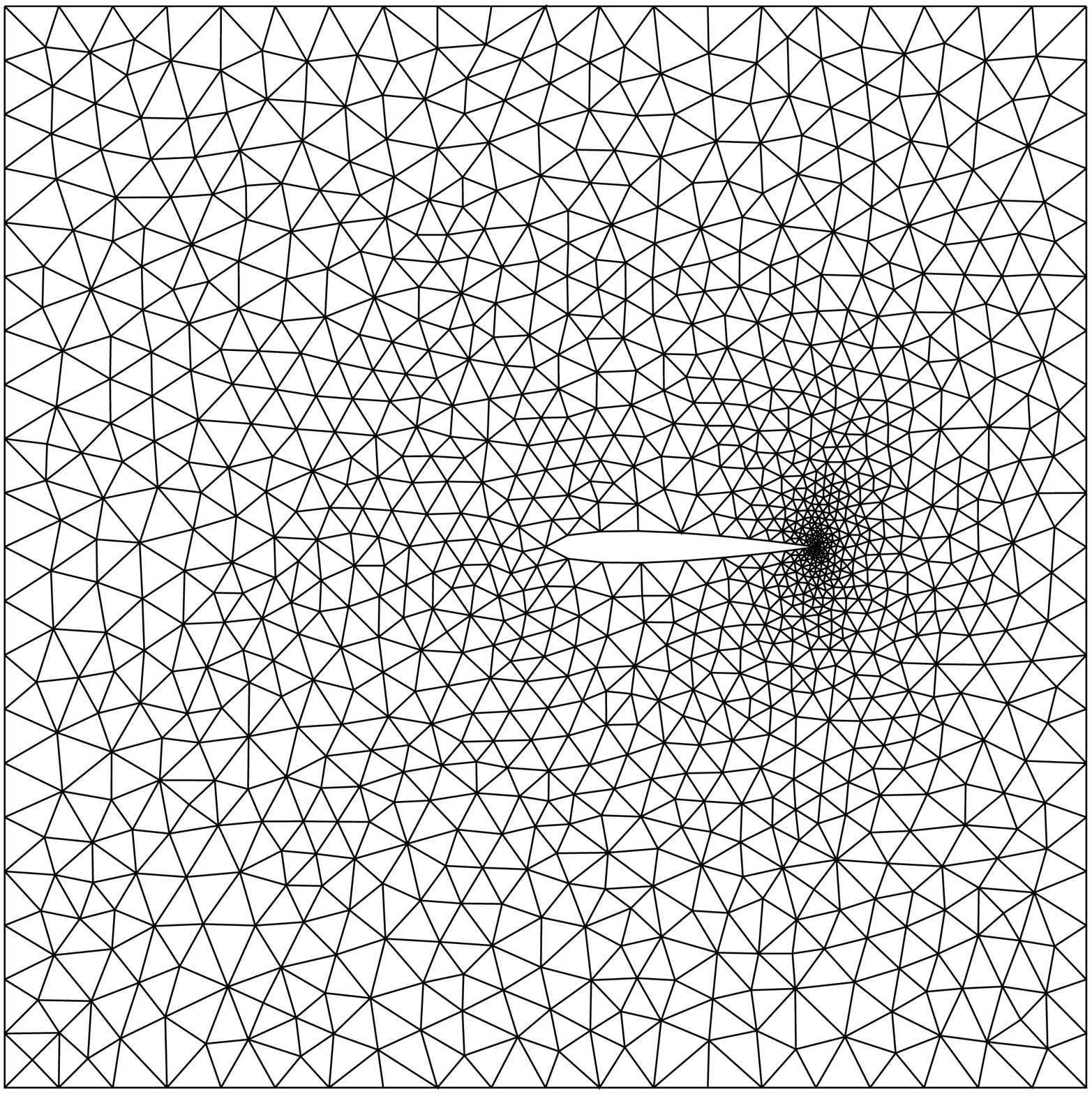}}}$
  \hspace*{.2in}
  $\vcenter{\hbox{\includegraphics[width=0.45\linewidth]{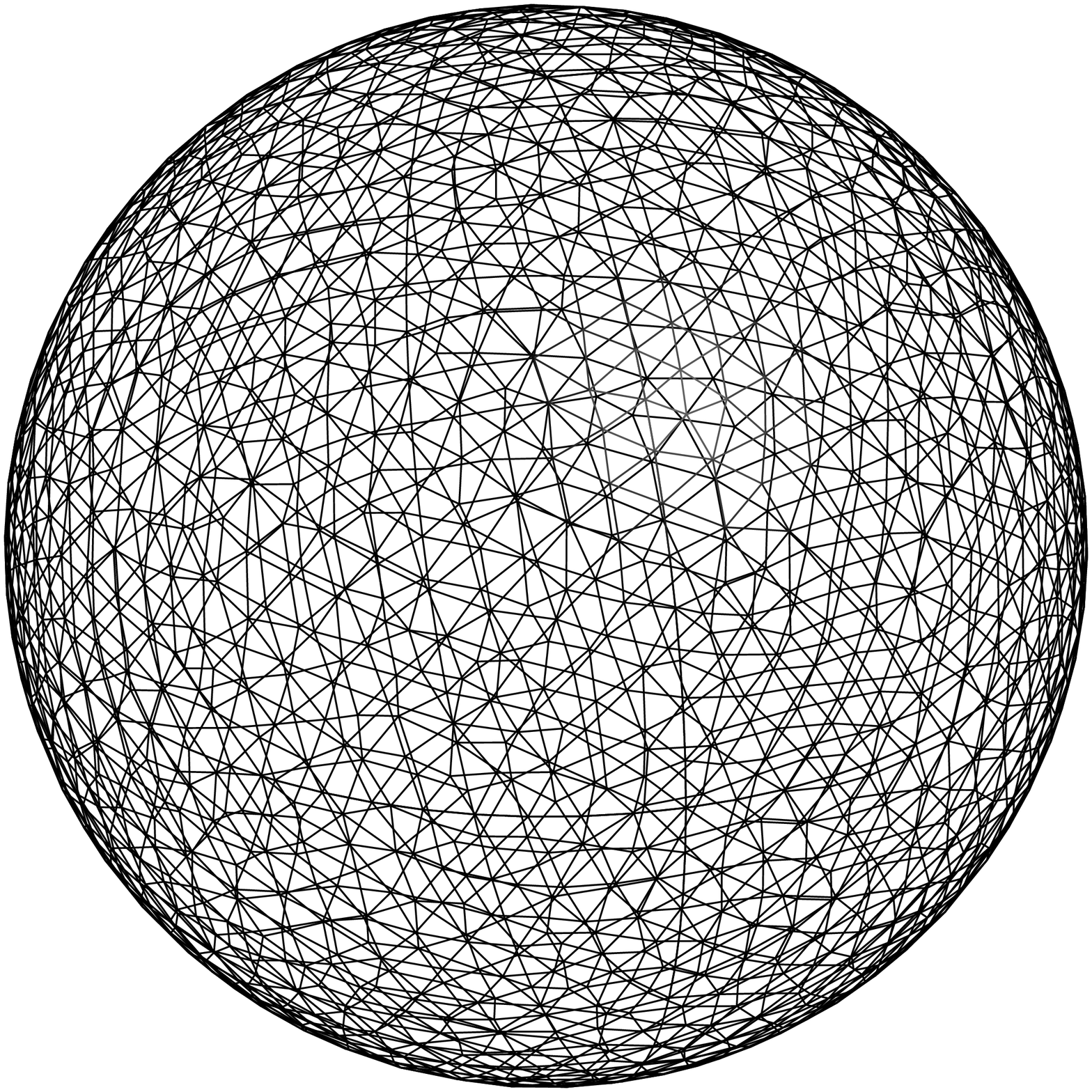}}}$
  \hspace*{.2in}
  $\vcenter{\hbox{\includegraphics[width=0.45\linewidth]{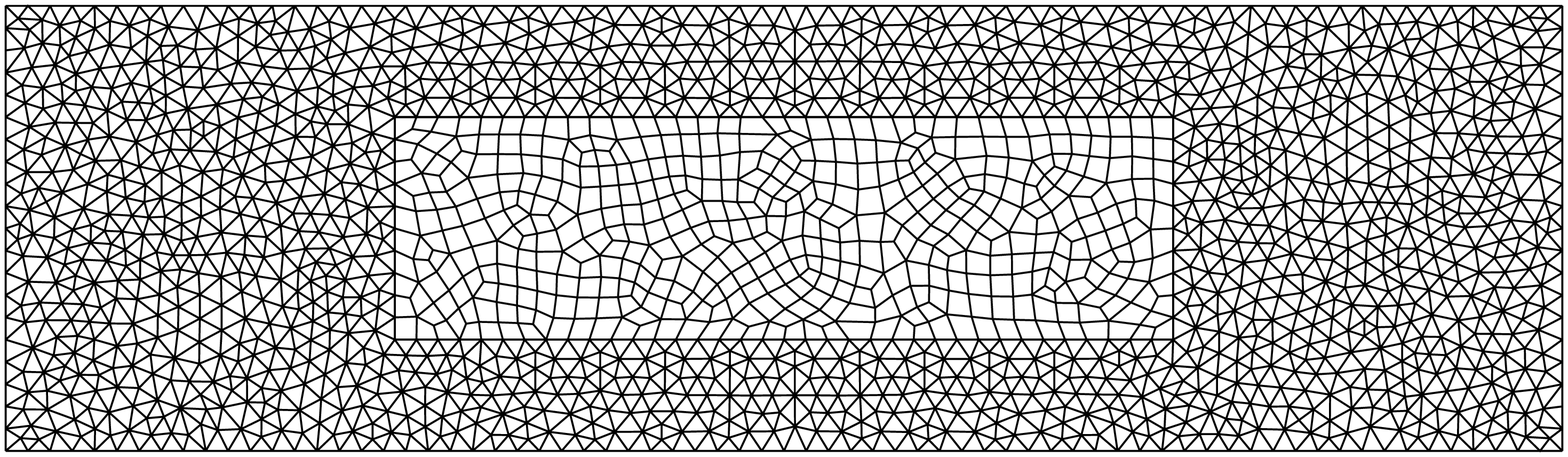}}}$
	\caption{Meshes clockwise from the top left: 3D prism A, two-dimensional NACA-0012 airfoil, rectangular domain of a hybrid hybrid tri/quad mesh, and spherical shell A.}
	\label{fig:meshes}
\end{figure}

For triangular and quadrilateral elements, the meshes are of simple rectangular domains.  We consider a sequence of meshes A-E, with A being the coarsest and E being the finest.  The meshes are not obtained through nested refinement, though the ratio of the number of elements in consecutive meshes is approximately 4.  Additionally, we discretize a spherical shell, i.e. a mesh without boundaries, and a domain containing the NACA-0012 airfoil i.e. a mesh with a cavity.  Finally, we color a hybrid quadrilateral/triangular element mesh.  Although there are triangles present in the mesh, we use four colors.  For tetrahedral elements, meshes A-C are prismatic domains along with a domain containing a three-dimensional extrusion of the NACA-0012 airfoil.  We plot some of the meshes considered in Figure \ref{fig:meshes}.

\begin{figure}
	\begin{tikzpicture}[scale=0.9]
	\begin{loglogaxis}[
	xlabel=No. sides,
	ylabel=No. conflicts,
	legend pos=north west
	]
	
	\addplot plot coordinates {
		(14775,     1697)
		(59547,    6631)
		(236372,    26097 )
		(942162,   104146)
		(3774165,   416542)
	};
	
	\addplot plot coordinates {
		(8888,     996)
		(35396,    3955)
		(141076,    16265 )
		(563108,   64913)
		(2249162,   260711)
	};
	
	\addplot plot coordinates {
		(14937,     1888)
		(102438,    12351)
		(720736,    81438 )
	};
	
	\logLogSlopeTriangle{0.6}{0.1}{0.4}{1}{black};
	\legend{Triangles\\ Quadrilaterals \\ Tetrahedra\\}
	
	\end{loglogaxis}
	
	\end{tikzpicture}
	\qquad
	\begin{tikzpicture}[scale=0.9]
	\begin{loglogaxis}[
	xlabel=No. sides,
	ylabel=Coloring time (s),
	legend pos=north west
	]
	
	\addplot plot coordinates {
		(14775,     0.52)
		(59547,    2.58)
		(236372,    12.52 )
		(942162,   65.19)
		(3774165,   326.05)
	};
	
	\addplot plot coordinates {
		(8888,   0.41)
		(35396,  2.48)
		(141076, 11.62)
		(563108, 58.42)
		(2249162,289.30)
	};
	
	\addplot plot coordinates {
		(14937,     0.23)
		(102438,    2.48)
		(720736,    25.70 )
	};
	
	\logLogSlopeTriangle{0.6}{0.1}{0.4}{1}{black};
	\legend{Triangles\\ Quadrilaterals \\ Tetrahedra\\}
	
	\end{loglogaxis}
	
	\end{tikzpicture}
	\caption{The number of conflicts vs. the number of sides in meshes A-E after the modified greedy coloring for different element geometries (left).  Coloring time for the same meshes (right).} \label{graphs}
\end{figure}
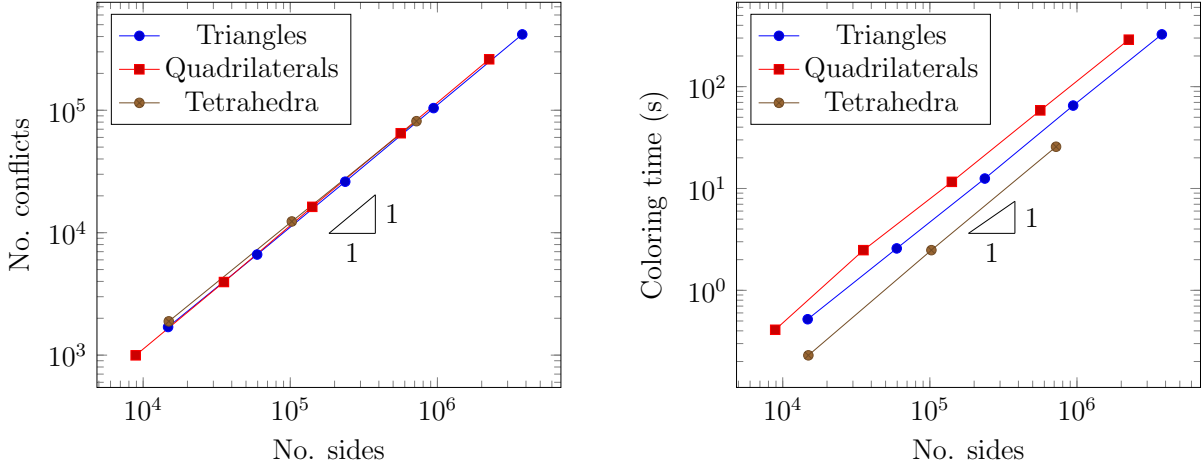

In Figure \ref{graphs}, we plot two metrics that measure the performance of the coloring algorithm.   On the left, we display the number of conflicts present in meshes A-E for the different element geometries (triangles, quadrilaterals, and tetrahedra) after the modified greedy coloring.  We note that the number of conflicts seems to scale linearly and depend only on the number of surfaces, i.e., edges or faces, in the mesh.  On the right of Figure  \ref{graphs}, we plot the time it takes these same meshes to be colored (both modified greedy coloring and conflict resolution) according to the two algorithms in Section \ref{sec:algorithm}.  The coloring runtime appears to scale linearly as well.

\begin{table}
	\centering
	\resizebox{\columnwidth}{!}{
		\begin{tabular}{|c|c|c|c|c|c|c|}
			\hline Triangular mesh & No. elements/sides  & \multicolumn{3}{c|}{No. edges in each color}& Coloring time (s)   \\ 
			\hline  A & 9,752/14,775 &  4,924 & 4,923 & 4,928  &  0.52 \\ 
			\hline  B & 39,502/59,547 &  19,845 & 19,845 & 19,857 & 2.58 \\ 
			\hline  C & 157,190/236,372 &  78,809 & 78,789  & 78,774 & 12.52 \\ 
			\hline  D & 627,326/942,162 &  314,069 & 314,058  & 314,035 & 65.19 \\ 
			\hline  E & 2,514,546/3,774,165 &  1,258,087 & 1,258,027  & 1,258,051 & 326.05 \\ 
			\hline NACA-0012 & 2,150/3,278 &  1,095 & 1,092 & 1,091 &  0.08 \\ 
			\hline Sphere A & 3,024/4,536 &  1,512 & 1,512 & 1,512 &  0.02 \\
			\hline Sphere B & 189,152/283,728 &  94,576 & 94,576 & 94,576 &  2.06 \\
			\hline
		\end{tabular} }
		
		\caption{Triangular mesh statistics; meshes A-E are on rectangular domains.  The NACA-0012 and spherical mesh A are displayed in Figure \ref{fig:meshes}.}
		\label{tab:tri}
	\end{table}
	\begin{table}

		\resizebox{\columnwidth}{!}{
			\begin{tabular}{|c|c|c|c|c|c|c|c|}
				\hline Quadrangular Mesh & No. elements/sides  & \multicolumn{4}{c|}{No. edges in each color}& Coloring time (s)   \\ 
				\hline  A & 4,373/8,888 & 2,222  & 2,230 & 2,221  & 2,215 & 0.41  \\ 
				\hline  B & 17,556/35,396 & 8,845  & 8,852 & 8,851 & 8,848 & 2.48 \\ 
				\hline  C & 70,255/141,076 & 35,279  & 35,271  & 35,256  & 35,270 & 11.62\\ 
				\hline  D & 280,988/563,108 & 140,794  & 140,762  & 140,779 & 140,773 & 58.42 \\ 
				\hline  E & 1,123,449/2,249,162 & 562,303  & 562,291 & 562,297 & 562,271 & 289.30  \\ 
				\hline  Hybrid tri/quad & 3,212/5,132 &  1,273 & 1,283  & 1,285 & 1,291 & 0.05 \\
				\hline
			\end{tabular} }
					\caption{Quadrangular and hybrid mesh statistics; meshes A-E are on rectangular domains.  The hybrid tri/quad domain is displayed in Figure \ref{fig:meshes}.}
					\label{tab:quad}		

		\end{table}
		\begin{table}

			\resizebox{\columnwidth}{!}{
				\begin{tabular}{|c|c|c|c|c|c|c|}
					\hline Tetrahedral Mesh & No. elements/sides  & \multicolumn{4}{c|}{No. edges in each color}& Coloring time (s)   \\ 
					\hline  A & 6,832/14,937 &  3,744 & 3,732  & 3,734 & 3,727 & 0.23 \\ 
					\hline  B & 48,879/102,438 &  25,681 & 25,681  & 25,577 & 25,596 & 2.48  \\ 
					\hline  C & 351,225/720,736 &  180,211  & 180,234 & 180,232 & 180,059 & 25.70\\ 
					\hline  NACA-0012 & 471,190/973,741 &  243,625 & 243,513 & 243,371 & 243,232 &  30.38 \\ 
					\hline
				\end{tabular} }
				\caption{Tetrahedral mesh statistics; meshes A-C are on prismatic domains.  Prismatic domain A is displayed in Figure \ref{fig:meshes}.  The NACA-0012 domain is a three-dimensional extrusion of the two-dimensional one in Figure \ref{fig:meshes}.}
				\label{tab:3D}

\end{table}

\subsection{Memory savings} \label{sec:memorysaved}
	By coloring the edges of the mesh, a substantial amount of memory can be saved.  If the edges were not colored, two buffers for the surface contribution to the left and right elements would have to be allocated.  The necessary memory for these buffers in double precision can be calculated with the formula $2  \times( N_p \text{ basis functions } )\times (N_{\text{eq}} \text{ equations })\times (N_{\text{s}} \text{ surfaces }) \times (8 \text{ bytes})$.  Two buffers are needed because there are two different surface contributions to the left and right elements of a surface in the DG method.  For the Euler equations, we show in Table \ref{tab:memory} the memory saved on mesh E of triangular elements for various orders of approximation $p$.  On GPUs, this is a non-negligible amount of memory, reaching 5 GBs for $p=5$.
	
		\begin{table}
		\centering

		\begin{tabular}{|c|c|c|c|c|c|}
				\hline Order of approximation $p$ & 1 & 2 & 3 & 4 & 5 \\ 
				\hline Memory saved (GB) & 0.72 & 1.44 & 2.41 & 3.62 & 5.07  \\
				\hline
		\end{tabular} 
		\caption{Memory saved from removing the buffer needed for the mesh E of 2,514,546/3,774,165 (elements/sides) of triangular elements for orders of approximation 1 to 5 for the Euler equations.} \label{tab:memory}
		\end{table}

\section{Data ordering}

On GPUs, optimal memory throughput is achieved when sequential threads access sequential locations in memory \cite{programmingguide}.  Coloring the edges with a minimal number of colors yields a straightforward procedure for ordering the elements and surfaces such that coalesced accesses are favoured.  We know that the elements in arbitrary unstructured grids cannot be ordered such that all memory accesses are coalesced.

The parallel DG-GPU flow solver described in \cite{giuliani} assigns one thread per edge in the surface computation kernel.  Each thread loads the data for left and right elements that share the edge.  These memory accesses for edges will be coalesced if sequential edges have sequential left and right elements, i.e., the $k$th thread loads data for the left element at memory location \texttt{offset1 + k} and data for the right element at memory location \texttt{offset2 + k}.  Without element and edge reordering, these accesses will likely be uncoalesced due to the irregular nature of unstructured meshes.

\begin{table}

	\begin{tabular}{c|c|c}
		Edge & Left Element & Right Element \\ 
		\cline{2-3} 0 & 54 & 32 \\ 
		\cline{2-3} 1 & 8 & 17 \\ 
		\cline{2-3} $\vdots$ & $\vdots$ & $\vdots$ \\ 
		\cline{2-3} $N_1-2$ & 25 & 12 \\ 
		\cline{2-3} $N_1-1$ & 2 & 13 \\ 
		
		\multicolumn{3}{c}{a)}
	\end{tabular} 
	\quad
	\begin{tabular}{c|c|c}
		Edge & Left Element & Right Element \\ 
		\cline{2-3} 0 & 0 & $N_1$ \\ 
		\cline{2-3} 1 & 1 & $N_1+1$ \\ 
		\cline{2-3} $\vdots$ & $\vdots$ & $\vdots$ \\ 
		\cline{2-3} $N_1-2$ & $N_1-2$ & $2N_1-2$ \\ 
		\cline{2-3} $N_1-1$ & $N_1-1$ & $2N_1-1$ \\
		
		\multicolumn{3}{c}{b)}
	\end{tabular} 
	\caption{a) arbitrary unstructured edgewise connectivity data for edges in color 1, where $N_1$ is the number of edges of color 1. b) the left and right elements of the edges of color 1 are renumbered such that they are consecutive in memory with respect to the ordering of edges in color 1.}
	\label{tab:connectivity}
\end{table}
\begin{table}

	\begin{tabular}{c|c|c}
		Edge & Left Element & Right Element \\ 
		\cline{2-3} $N_1$ & 103 & 65 \\ 
		\cline{2-3} $N_1 + 1$ & 0 & 43 \\ 
		\cline{2-3} $\vdots$ & $\vdots$ & $\vdots$ \\ 
		\cline{2-3} $N_2-2$ & 34 & 47 \\ 
		\cline{2-3} $N_2 -1$ & 2 & 209 \\
		
		\multicolumn{3}{c}{a)}
	\end{tabular} 
	\quad
	\begin{tabular}{c|c|c}
		Edge & Left Element & Right Element \\ 
		\cline{2-3} $N_1$ & 0 & 43 \\ 
		\cline{2-3} $N_1 + 1$ & 2 & 209 \\ 
		\cline{2-3} $\vdots$ & $\vdots$ & $\vdots$ \\ 
		\cline{2-3} $N_2-2$ & 102 & 31 \\ 
		\cline{2-3} $N_2 - 1$ & 103 & 65 \\
		
		\multicolumn{3}{c}{b)}
	\end{tabular} 
	
		\caption{a) connectivity data for edges in color 2, where $N_2$ is the number of edges of color 2. b) edges are reordered with ascending left element.}
		\label{tab:color2}
	
\end{table}

	\begin{table}

		\resizebox{\columnwidth}{!}{
			\begin{tabular}{|c|c|c|c|c|}
				\hline 5 colors & 3 colors &  Element renumbering  & Edge reordering & Solver runtime (s) \\ 
				\hline x &  &  &  & 166.79 \\
				\hline  & x &  &  & 158.14 (5.1)\\
				\hline x &  & x &  & 177.93 \\
				\hline  & x & x &  & 153.31 (8.0) \\
				\hline x &  & x & x & 144.75 (13.2) \\
				\hline  & x & x & x & 129.34 (22.4) \\
				\hline
			\end{tabular} }
				\caption{Execution runtimes for 1000 timesteps of the double Mach test problem with $p=1$ on mesh D presented in Table \ref{tab:tri} for different data reorganization strategies.  The number in parentheses is the percent speed-up relative to the unordered, unrenumbered mesh of five colors.}
				\label{tab:timing}
		\end{table}

	Mesh connectivity is stored as pointers: an edge points to the two elements that share it.  Table \ref{tab:connectivity} a) shows an excerpt of edgewise connectivity data required for an unstructured mesh.  The $k$th row stores data for edge $e_k$.  The first and second columns store the element numbers of each edge's left and right element.  For example, $e_0$ has left element $\Omega_{54}$ and right element $\Omega_{32}$.
	
	We propose the following element and edge renumbering method to reduce occurrences of irregular access patterns.  This is done as a mesh preprocessing stage.  First, color the edges of the mesh and reorder them based on their assigned color, i.e. edges 0 to $N_1$ are of color 1, where $N_1$ is the number of edges of color 1, followed by edges of color 2 and 3.  That is, we have a connectivity table as shown in Table \ref{tab:connectivity} a) for color 1.
	
	Next, the left and right elements of the $k$th edge in color 1 are renumbered as element $k$, and $N_1+k$, respectively, where $N_1$ is the number of edges in color 1.  Table \ref{tab:connectivity} b) represents the renumbered connectivity data for the edges in color 1.  Because every element has exactly one edge of color 1, all the elements in the mesh are renumbered.
	
	This procedure yields entirely coalesced acccesses for the edges in color 1, i.e. for approximately a third of the edges for triangles and a fourth of the edges for quadrangles and tetrahedra. 
	
	Because all the elements have been renumbered, the order of the left and right elements of edges in the remaining color groups may not be consecutive, as shown in Table \ref{tab:color2} a).  We can no longer renumber elements, but we can reorder edges in these groups so that the left element numbers are in ascending order, see Table \ref{tab:color2} b).  Therefore some memory transfers for left elements will be coalesced.  
	
	Reorganizing the mesh can be done efficiently using hash tables even for problems of substantial size.  For example, the total element reordering and edge renumbering time of mesh D with three colors presented in Table \ref{tab:tri} (627,326 elements, and 942,162 sides) is 7.98 seconds.  For the other meshes of triangular elements (meshes A-E), see Table \ref{tab:renum}.
	
	\begin{table}
	\centering

	\begin{tabular}{|c|c|c|c|c|c|}
			\hline Mesh & A & B & C & D & E \\ 
			\hline Time (s) & 0.08 & 0.37 & 1.81 & 7.98 & 33.26 \\ 
			\hline
	\end{tabular} 
	\caption{Total renumbering and reordering times for the two-dimensional meshes A-E of triangles with three colors.} \label{tab:renum}
	\end{table}

	\subsection{Data ordering examples}
	We now apply this element and edge renumbering scheme to the DG-GPU solver for the double Mach reflection test problem described in \cite{giuliani} on an NVIDIA GTX 580 graphics processing unit, which has 3 GB of video memory.  We run the solver with a polynomial approximation $p = 1$ for 1000 timesteps on mesh D with triangular elements referenced in Table \ref{tab:tri}.  In Table \ref{tab:timing}, we measure the solver performance by comparing the runtimes of a mesh colored with an optimal three colors and meshes with a suboptimal five colors.  Compared with a mesh of five colors without element renumbering or edge reordering, we find that reducing the number of colors gives minor impovement ($\sim$ 5 \% speed-up).  These savings result from less overhead from kernel launches \cite{luo}.  We also report the runtime for element-renumbered meshes that were colored with three and five colors.  Here, using more colors than required actually increases runtime.  Next, we combine element renumbering and edge reordering. With five colors, we have an improvement of $\sim$ 13 \%.  The speed-up increases to approximately 22 \% when the renumbered and reordered mesh has three colors.

	\section{Conclusion}
We have presented a simple method to mitigate the race condition that arises during the surface integral evaluation of shared-memory-based parallel DG and finite volume type flow solvers.  Although we do not test on a finite volume solver, our approach is applicable to this method as well.  The main motivation of this work was to reduce the memory footprint of the code by eliminating buffers \cite{giuliani, nodaldg, corrigan}.  This procedure colors the edges of unstructured triangular, tetrahedral and quadrilateral meshes using an optimal number of colors; as a result, the number of kernel launches is reduced, contributing to code simplicity.  We do not have a proof that the coloring algorithm always terminates.  However, we have not encountered a mesh for which it failed to do so.  We note that practical edge coloring algorithms existing in literature \cite{complexcolors, fiol2012} do not have formal proofs either.

We have proposed a coloring algorithm extension to adaptive mesh refinement.  The original coloring can be done in a preprocessing stage and adaptive coloring is a simple mapping from the original coloring, which can be done quickly and in parallel.  

The same coloring is used for reorganizing the DOFs in memory to reduce access latencies.  The reordering scheme allows for approximately a 22 \% speed-up when applied to a DG solver of the Euler equations in two-dimensions.  

The edge coloring algorithm can also serve as an approach to quadrilateral mesh generation.  Deleting all the edges of one chosen color creates a mesh comprised of quadrilaterals everywhere in the domain interior.  This is similar to Blossom-Quad \cite{quadblossom}, a perfect matching algorithm.   It will be interesting to investigate if our coloring algorithm yields meshes of comparable quality in a similar amount of time.  We leave this for a future work.

\section{Acknowledgements}
This work was supported in part by the Natural Sciences and Engineering Research Council of Canada grant 341373-07, and Alexander Graham Bell PGS-D grant.

	\bibliography{mybib}{}
	\bibliographystyle{ieeetr}
\end{document}